\documentclass[twocolumn,prl,showpacs]{revtex4}
\usepackage{graphics}
\usepackage{graphicx}
\usepackage{amsmath}
\usepackage{amssymb}
\usepackage{subfigure}
\usepackage{longtable}
\usepackage{multirow}
\usepackage{bm}
\usepackage[utf-8]{inputenc}

\newcommand{\ve}[1]{\bm{\mathrm{#1}}}

\graphicspath{{.}}

\begin{document}
\title{Graphene on metals: a Van der Waals density functional study}
\author{M. Vanin}
\affiliation{Center for Atomic-scale Materials Design, Department of
Physics \\ Technical University of Denmark, DK - 2800 Kgs. Lyngby, Denmark}
\author{J. J. Mortensen}
\affiliation{Center for Atomic-scale Materials Design, Department of
Physics \\ Technical University of Denmark, DK - 2800 Kgs. Lyngby, Denmark}
\author{A. K. Kelkkanen}
\affiliation{Center for Atomic-scale Materials Design, Department of
Physics \\ Technical University of Denmark, DK - 2800 Kgs. Lyngby,
Denmark}
\author{J. M. Garcia-Lastra}
\affiliation{Center for Atomic-scale Materials Design, Department of
Physics \\ Technical University of Denmark, DK - 2800 Kgs. Lyngby,
Denmark}
\author{K. S. Thygesen}
\affiliation{Center for Atomic-scale Materials Design, Department of
Physics \\ Technical University of Denmark, DK - 2800 Kgs. Lyngby,
Denmark}
\author{K. W. Jacobsen}
\affiliation{Center for Atomic-scale Materials Design, Department of
Physics \\ Technical University of Denmark, DK - 2800 Kgs. Lyngby, Denmark}

\date{\today}

\begin{abstract}
  We use density functional theory (DFT) with a recently developed van
  der Waals density functional (vdW-DF) to study the adsorption of
  graphene on Al, Cu, Ag, Au, Pt, Pd, Co and Ni(111) surfaces. In
  constrast to the local density approximation (LDA) which predicts
  relatively strong binding for Ni,Co and Pd, the vdW-DF predicts weak
  binding for all metals and metal-graphene distances in the range
  3.40-3.72 \AA. At these distances the graphene bandstructure as
  calculated with DFT and the many-body G$_0$W$_0$ method is basically
  unaffected by the substrate, in particular there is no opening of a
  band gap at the $K$-point.
\end{abstract}

\maketitle

The recently reported synthesis of graphene\cite{Novoselov:2004}, a
single layer of graphite, ontop of a SiO$_2$ substrate has renewed the
interest for this unique material. The uniqueness of this 2D crystal
is mainly due to its very peculiar band structure, with the $\pi$ and
$\pi^*$ bands showing linear dispersion around the Fermi level where
they touch in a single point.  The great variety of physics and
chemistry which derives from this electronic structure makes graphene
very attractive for a range of applications. In particular, its high
stability and good conductivity under ambient conditions makes it an
interesting candidate for future nano-scale
electronics\cite{Geim:2007}. In this perspective, the interaction of
graphene with metallic contacts plays a fundamental role.  Moreover,
catalytic growth of graphene on transition metal surfaces from carbon
containing gases has become a standard way to obtain high quality
graphene
samples\cite{Wintterlin:2009,Kim:2009,Coraux:2009,Sutter:2008}.
Nevertheless the nature of the metal-graphene chemical bond is still
not well understood.\cite{Wintterlin:2009}.

The widely used density functional theory (DFT) with local and
semi-local functionals for exchange and correlation usually provides
an accurate description of covalent and ionic chemical bonds. On the
other hand it fails to reproduce non-local dispersive forces, in
particular van der Waals forces, which are important in weakly bonded
materials such as graphite, molecular crystals, and many organic
compounds.\cite{Hobza:1995,Kristyan:1994,Ruiz:1995}. It is also well
known that the local density approximation (LDA) tends to overbind
systems where van der Waals interactions are important, while the
generalized gradient approximations (GGA) usually tend to
underestimate the binding in these systems. In the case of graphene on
metals many GGAs, contrary to experiments, predicts no binding at all,
and therefore most theoretical work on graphene-metal interfaces has
relied on the LDA.  In view of the fact that LDA in general cannot be
considered a reliable approximation in non-homogeneous systems such as
surfaces and molecules, the graphene-metal interface clearly calls for
new and improved functionals.

The interaction of graphene with the (111) surfaces of Al, Cu, Ag, Au,
Pt, Pd, Co and Ni was studied in Ref.\cite{Giovannetti:2008} using the
LDA approximation. The LDA results divide the metals into two classes:
Co, Ni and Pd which bind graphene strongly and Al, Cu, Ag, Au and Pt
which bind graphene weakly. In contrast PBE\cite{PBE:1996} gives no
binding of graphene at room temperature\cite{Fuentes:2008}.  This
remarkable disagreement between the two most commonly used
approximations of DFT might be related to the incorrect description of
dispersion interactions in both of the functionals.

In this paper we use the recently developed van der Waals density
functional (vdw-DF)\cite{Dion:2004,Dion:2005} to investigate the nature of the
bonding at the metal-graphene interface. The functional is explicitly constructed to include
non-local dispersion interactions and has proven successful in several cases
where standard functionals fail, such as rare gases\cite{Dion:2004},
benzene dimers\cite{Puzder:2006,Thonhauser:2006}, graphite
\cite{Ziambaras:2007}, polymers\cite{Kleis:2007},
DNA\cite{Cooper:2008} and organic molecules on
surfaces\cite{Romaner:2009,Moses:2009,Langreth:2009}. Within the
vdw-DF approximation, the exchange-correlation energy is
\begin{equation}
E_{xc}^{\mathrm{vdw-DF}}=E_x^{\mathrm{revPBE}}+E_c^{\mathrm{LDA}}+E_c^{\mathrm{nl}}
\end{equation}
where $E_x^{\mathrm{revPBE}}$ is the revPBE\cite{Zhang:1998} exchange
energy, $E_c^{\mathrm{LDA}}$ is the LDA correlation energy and
$E_c^{\mathrm{nl}}$ is the non-local correction given by
\begin{equation}
  \label{eq:E_nl}
  E_c^{\mathrm{nl}}=\frac{1}{2} \iint n(\mathbf{r_1}) n(\mathbf{r_2})
  \phi(q_1,q_2,r_{12}) \mathrm d \ve r_1 \mathrm d \ve r_2
\end{equation}
where $r_{12}=|\ve r_1 - \ve r_2|$ and $q_1$ and $q_2$ are values of a
universal function $q_0(n(\ve r), |\nabla n(\ve r)|)$.  Eq.
\eqref{eq:E_nl} is efficiently evaluated by factorizing the
integration kernel $\phi$ and by using fast Fourier transform to
compute the selfconsistent potential as proposed in Ref.\
\cite{Soler:2009} and implemented in the real-space projector
augmented wave GPAW code\cite{Mortensen:2005}.

In this study we consider Al, Cu, Ag, Au, Pt, Ni, Co and Pd metal
(111) surfaces. We fix the atoms in the metal slabs at their
experimental lattice parameters and relax the graphene sheet using the
vdw-DF Hellmann-Feynman forces. We use a (6,6,1) and (4,4,1) Monkhorst
Pack k-point sampling respectively for the smaller (Ni, Cu, Co) and
the larger (Au, Ag, Pt, Pd, Al) orthorombic unit cells. The metal
slabs are modeled with $4$ atomic layers and a vacuum of $14\,
\rm{\AA}$ in the direction normal to the surface; the grid spacing is
$0.16 \, \rm{\AA}$. The calculated binding energies and distances for
the relaxed structures are listed in Table \ref{tab:table1}. The
vdW-DF results show that the metal-graphene interaction is relatively
similar across the different metals. This is in contrast with the LDA
prediction of two separate classes of metal-graphene interfaces, as
found in very good agreement with \cite{Giovannetti:2008}. We also
repeated the same calculations using the revPBE functional and we
obtained no binding for any of the metals. Interestingly, for the
systems that LDA finds to be weakly bonded (Al, Cu, Ag, Au and Pt),
the binding energies obtained with the vdw-DF are very similar to the
LDA ones.  Nevertheless the binding distances are systematically
slightly larger in the vdw-DF case.  In fact it has been reported that
the vdw-DF functional usually produces equilibrium distances somewhat
larger than experiments\cite{Langreth:2009}.  In the case of Ni, Co
and Pd, on the other hand, the relatively strong binding predicted by
LDA is not found by the vdw-DF functional.

\begingroup
\squeezetable
\begin{table}
  \begin{tabular}{c|c|c|c|c|c|c|c|c|c}
    \hline
    \hline
    &&Al&Cu&Ag&Au&Ni&Co&Pt&Pd\\
    \cline{1-8}
    \hline
    \multirow{2}{*}{vdw-DF}
    &$d$ (\AA)&3.72&3.58&3.55&3.57&3.50&3.40&3.67&3.50\\  
    &$E_b$ (meV)&35&38&33&38&37&30&43&39\\
    &$\Delta E_{\rm{F}}$ (eV)&-0.51&-0.43&-0.40&+0.21&+0.13&-0.20&+0.66&+0.65\\  
    &$\delta Q \, (10^{-3}e)$&-8.0&-4.0&-5.0&+0.4&-3.0&-5.0&+5.0&+0.5\\  
    \hline   
    \multirow{2}{*}{LDA}
    &$d$ (\AA)&3.46&3.21&3.32&3.35&2.08&2.08&3.25&2.33\\  
    &$E_b$ (meV)&25&35&45&31&123&175&33&79\\
    \hline    
    \multirow{2}{*}{revPBE}
    &$d$ (\AA)&-&-&-&-&-&-&-&-\\  
    &$E_b$ (meV)&-&-&-&-&-&-&-&-\\
    \hline
    \hline
  \end{tabular}
  \caption{Binding energies ($E_b$) per carbon atom and binding 
    distances ($d$) of graphene on metal (111) surfaces; '-' means no binding.
    Fermi level shift $\Delta E_{\rm{F}}$ and charge transfer
    $\delta Q$ of graphene adsorbed on the different metals at the 
    vdw-DF equilibrium separation. Negative (positive) $\Delta E_{\rm{F}}$
    indicates $n$ ($p$)-type doping. Negative (positive) $\delta Q$
    indicates electron transfer to (from) the graphene layer. The
    charge transfer has been evaluated according to the Bader
    scheme\cite{Tang:2009}.}
  \label{tab:table1}
\end{table}
\endgroup

\begin{figure}
    \includegraphics[width=0.45\textwidth]{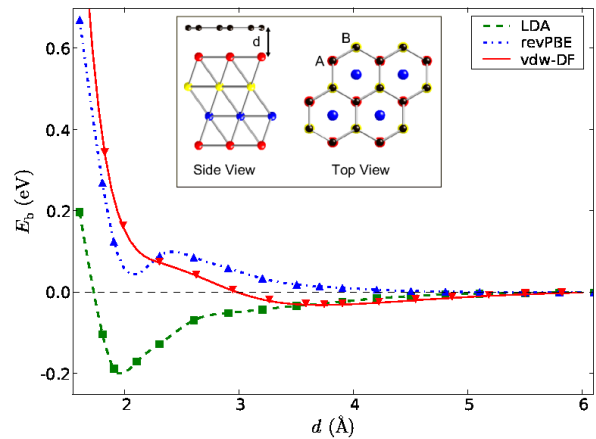}
    \caption{Binding energy ($E_b$) per carbon atom of graphene on
      the Ni(111) surface calculated with LDA, revPBE and vdw-DF
      functionals. The graphene is adsorbed in the top-fcc
      configuration.}
  \label{fig:fig1}
\end{figure}

In order to analyze these results, we now focus on the interaction
between graphene and Ni(111). Fig.\ \ref{fig:fig1} shows the
binding curves for graphene on the Ni(111) surface calculated with the
LDA, revPBE and vdw-DF functionals. The revPBE curve is positive at
all distances, while the LDA curve shows a relatively deep minimum at
$\sim 2 \, \rm{\AA}$ consistent with previous LDA calculations. The
vdw-DF result lies in between, following the revPBE curve at small
separations and the LDA curve at larger separations, and it predicts a
shallow minimum at $3.5 \, \rm{\AA}$.  Note that a local minimum is
found by the revPBE functional around $2 \, \rm{\AA}$.

\begin{figure*}
  \textup{Graphene on Ni(111)
}

  \subfigure{
    \includegraphics[width=0.40\textwidth]{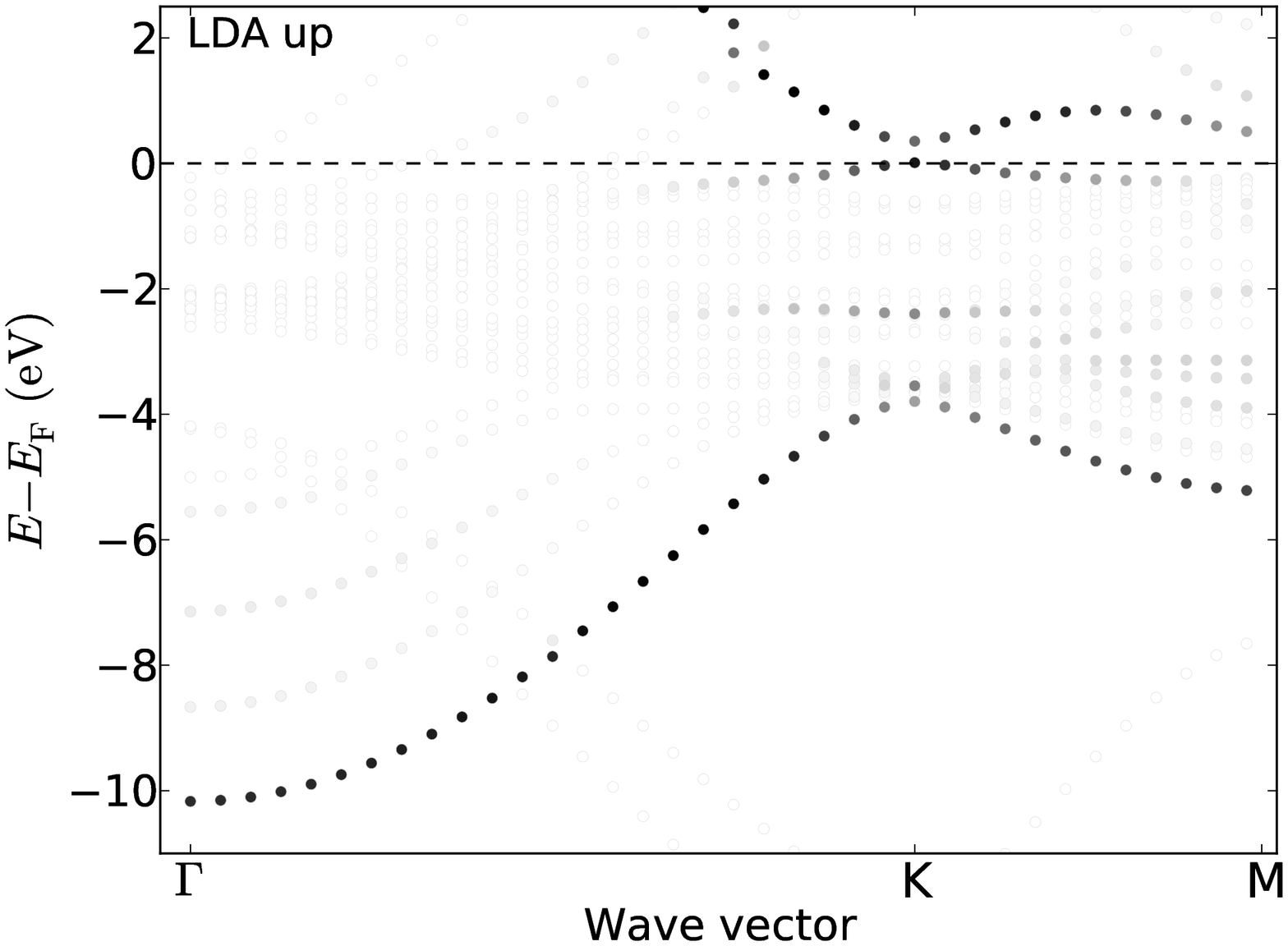}
    \label{fig:subfig1}
  } 
  \subfigure{
    \includegraphics[width=0.40\textwidth]{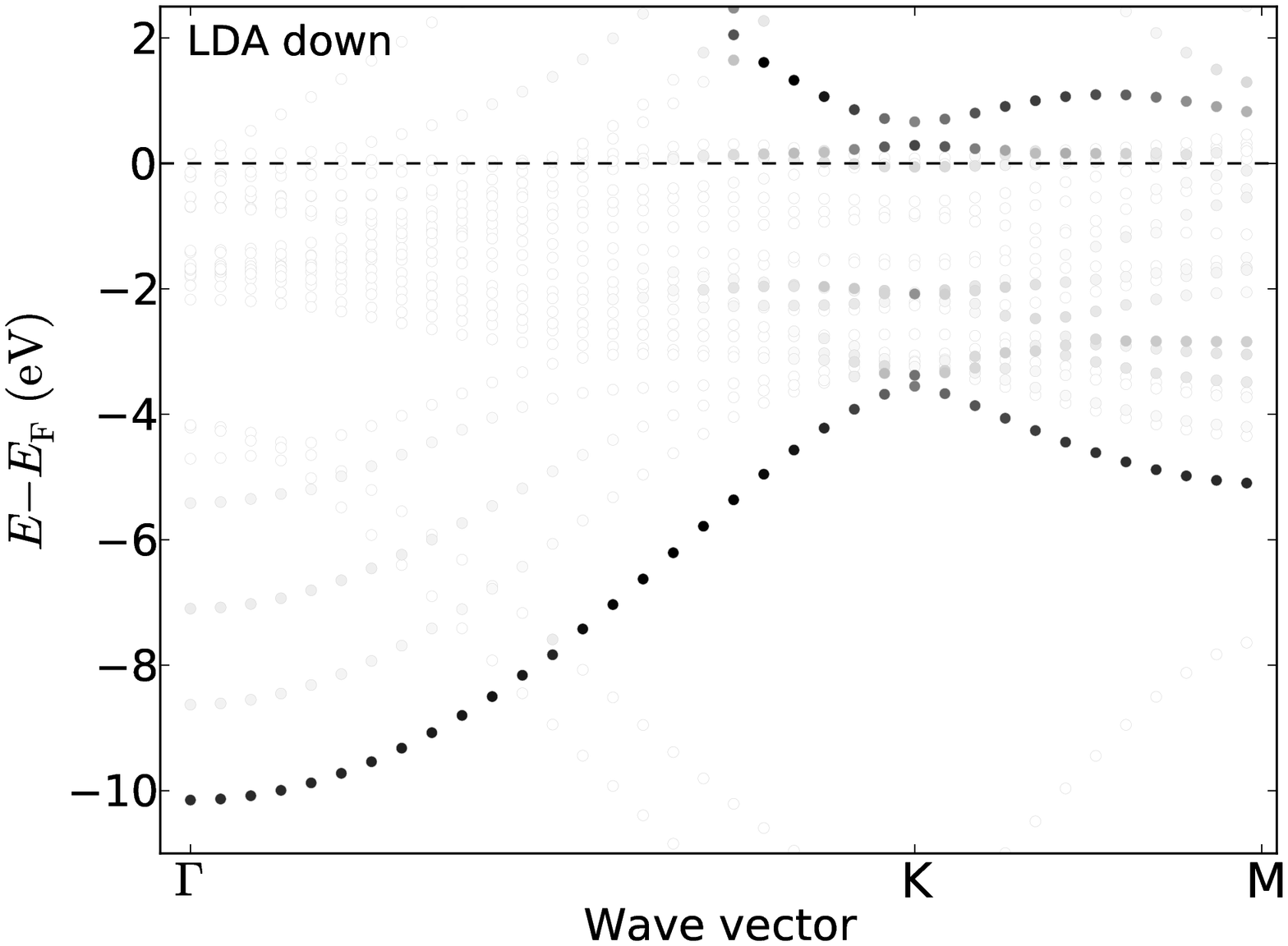}
    \label{fig:subfig2}
  } 
  \subfigure{
    \includegraphics[width=0.40\textwidth]{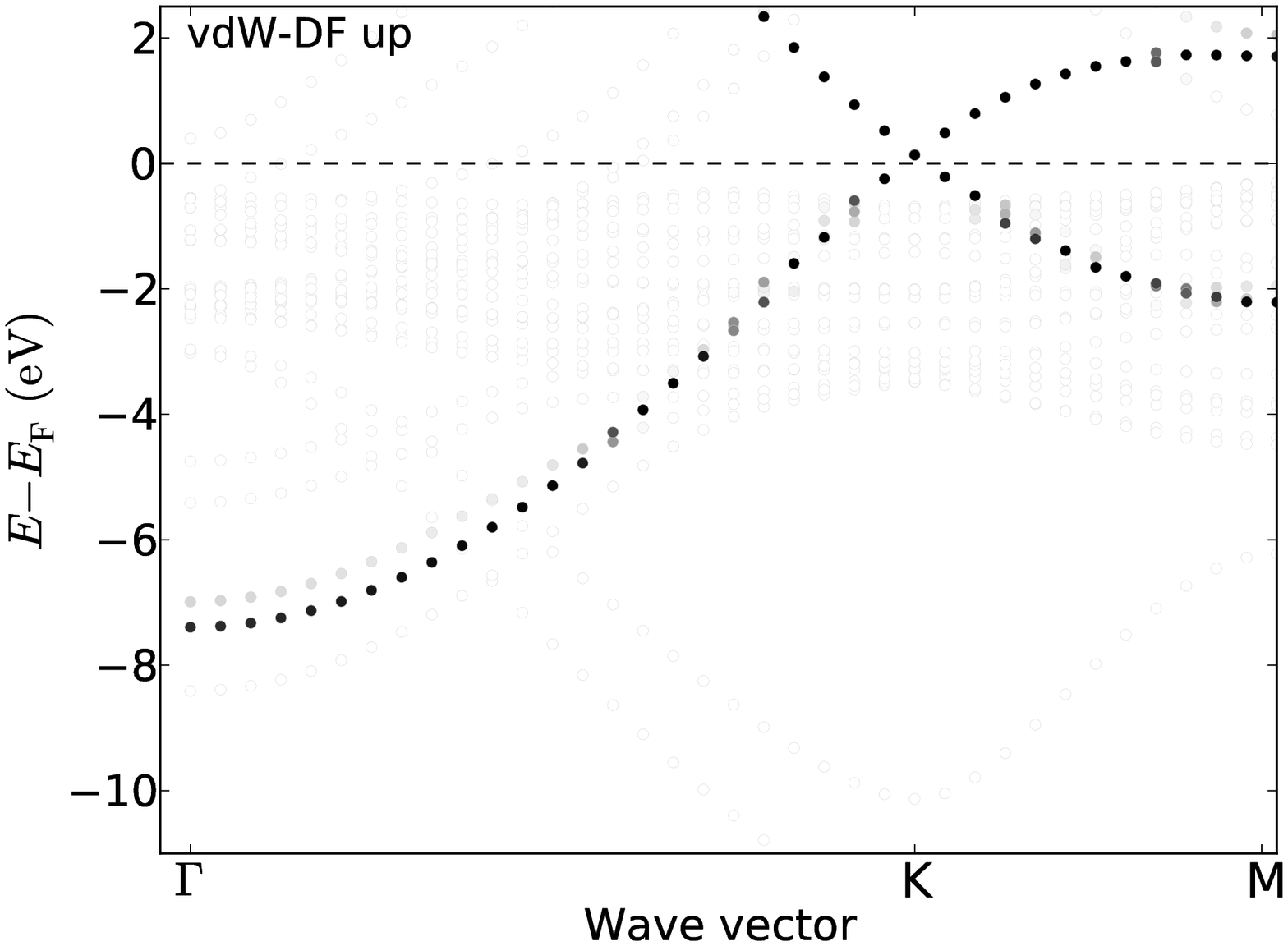}
    \label{fig:subfig3}
  }
  \subfigure{
    \includegraphics[width=0.40\textwidth]{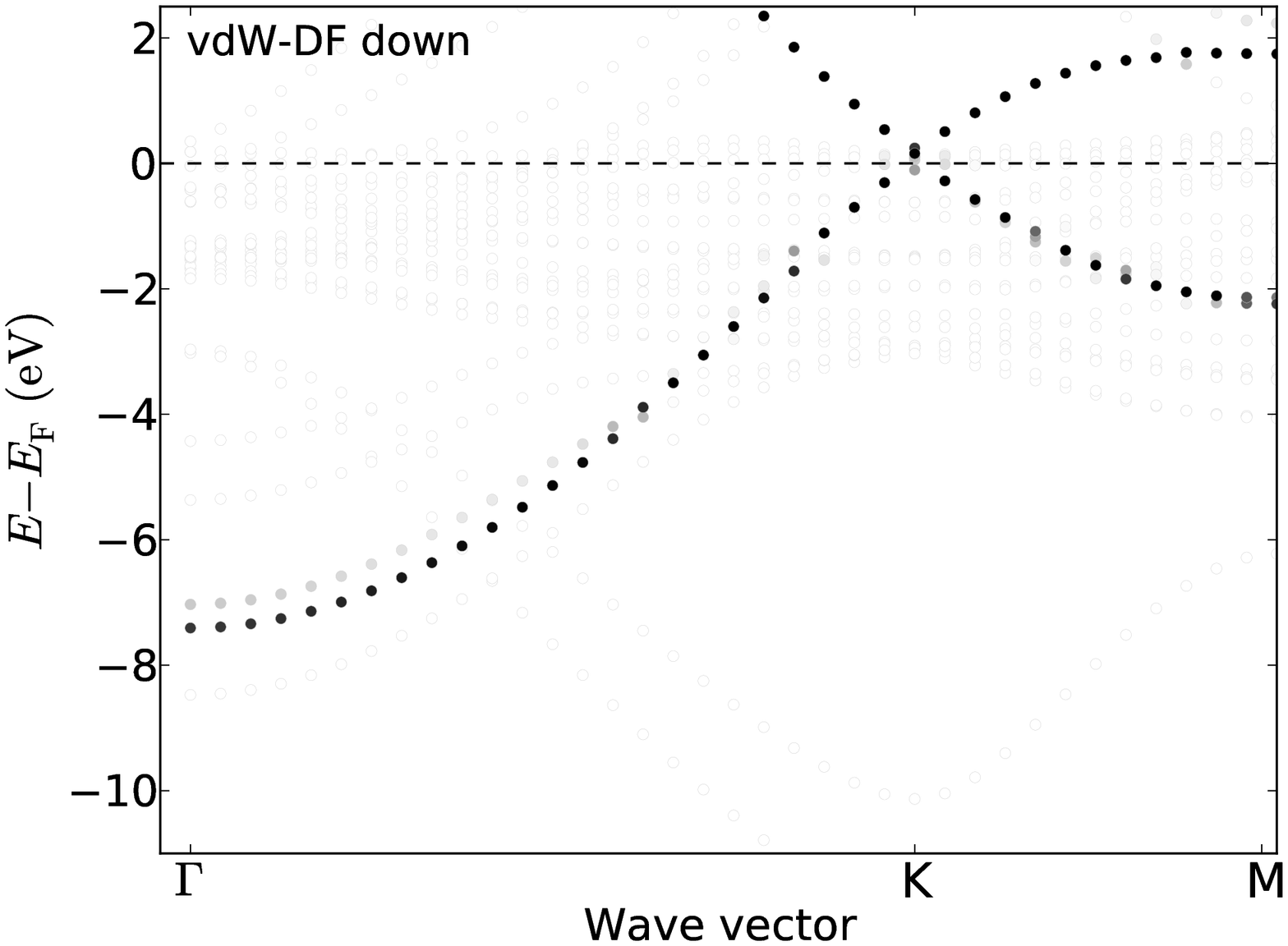}
    \label{fig:subfig4}
  }
  \caption{LDA (top) and vdw-DF (bottom) spin polarized band
    structures for graphene on Ni(111) in the top-fcc configuration.
    Darker dots represent larger weight of the carbon $p_z$ orbitals.}
  \label{fig:Ni_bands}
\end{figure*}

\begin{figure*}[]
  \subfigure{
    \includegraphics[width=0.45\textwidth]{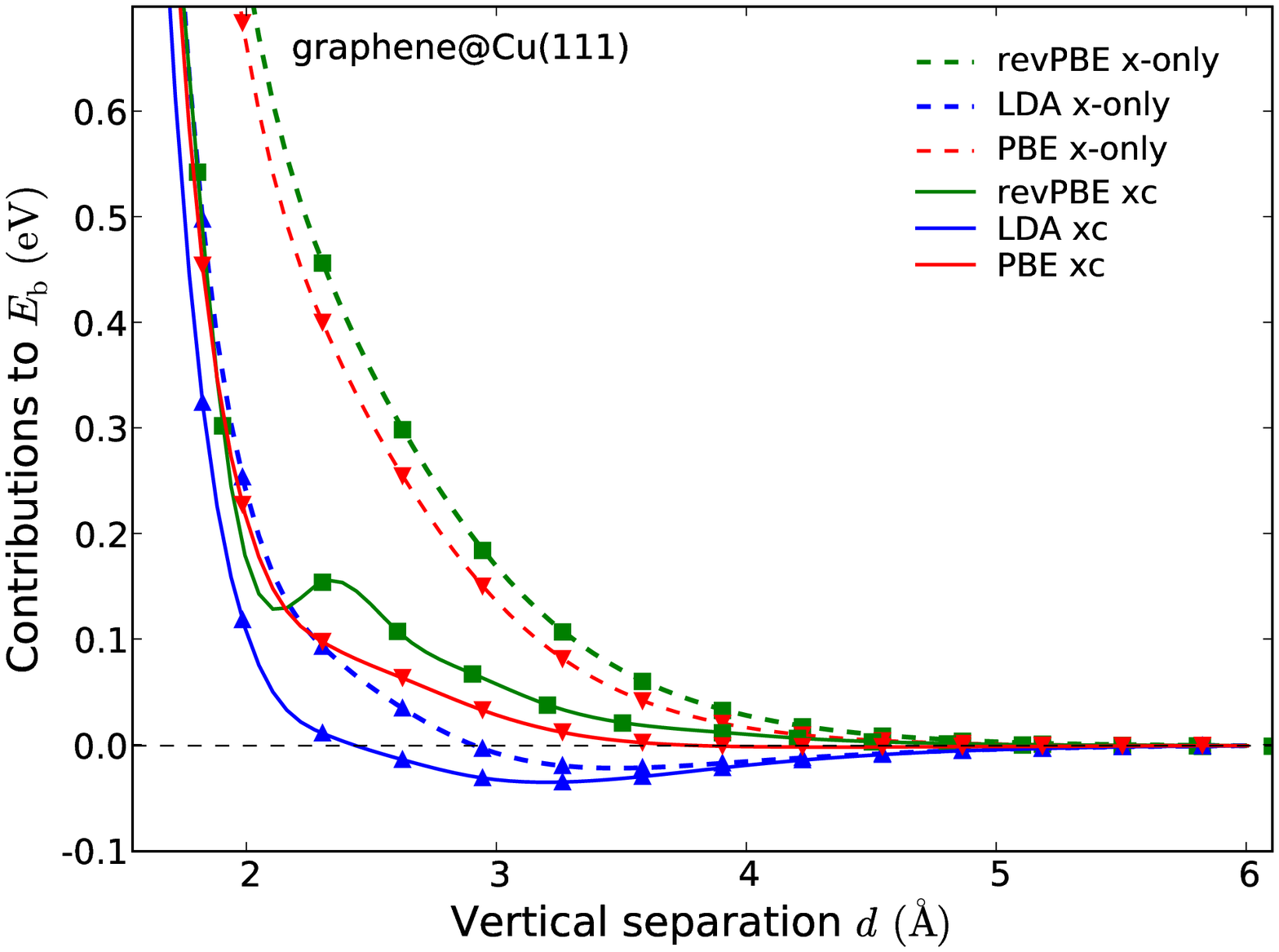}
    \label{fig:Ni_x-only}
  } 
  \subfigure{
    \includegraphics[width=0.45\textwidth]{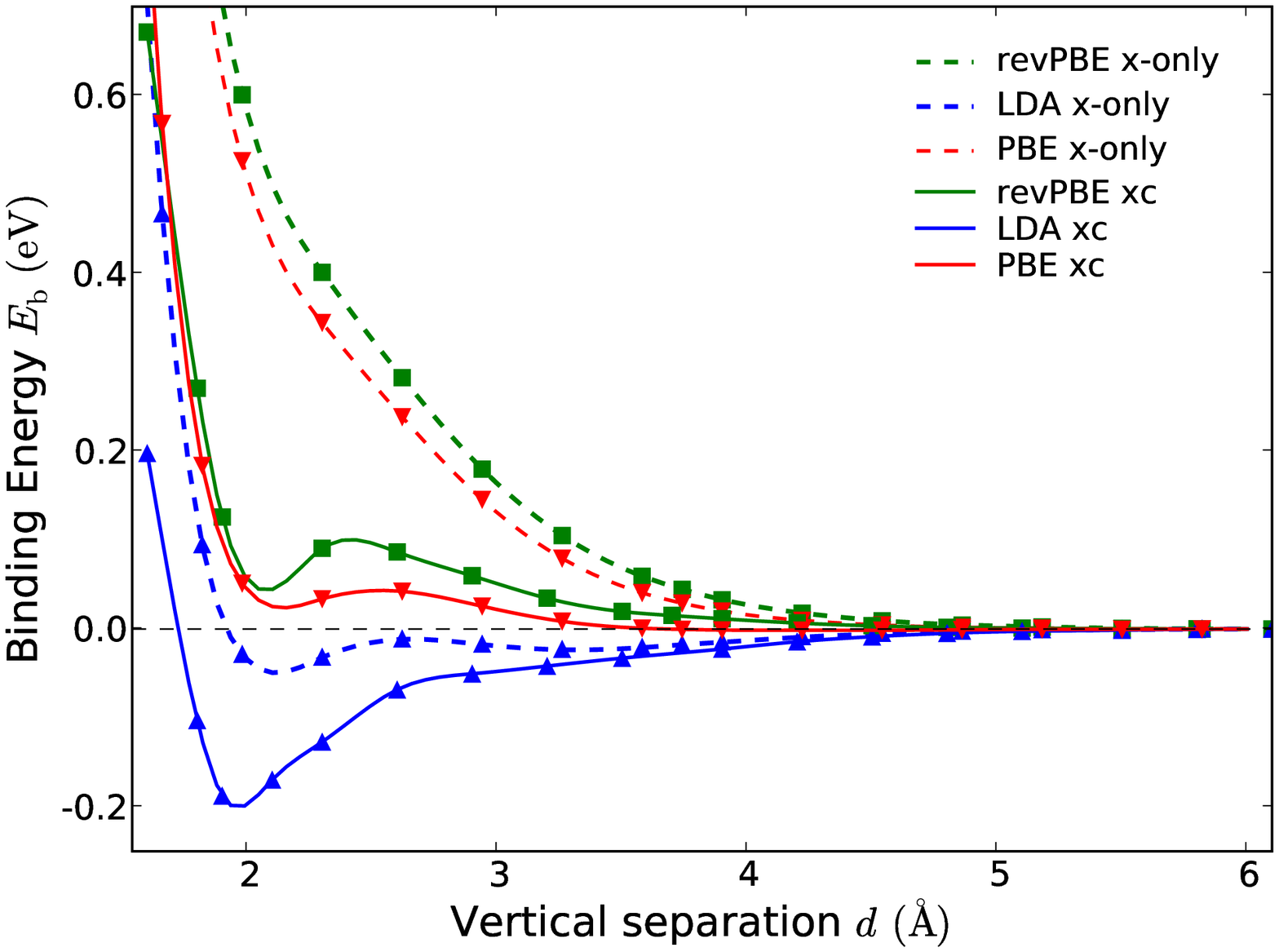}
    \label{fig:Cu_x-only}
  } 
  \caption{Decomposition of the binding energy $E_b$ into
    exchange-only contributions (dashed lines - only the correlation
    term is removed) and total binding energy (full lines) for
    different functionals.}
  \label{fig:binding_contributions}
\end{figure*}

In Fig.\ \ref{fig:Ni_bands} we show the calculated bandstructure of
graphene on Ni(111). The color of the dots indicate the weight of the
corresponding Bloch eigenstate on the carbon $p_z$ orbitals with
darker meaning larger weight. 
In free graphene, the carbon $p_z$ orbitals placed at A sites
($p_z^{\mathrm{A}}$) are decoupled from the $p_z$ orbitals at B sites
($p_z^{\mathrm{B}}$) at the Dirac point, thus producing two degenerate
states (see inset in Fig. \ref{fig:fig1} for the structure).
Since the A sites are located directly on top of Ni atoms at a close
distance in the LDA calculation (2.08 \AA), a strong hybridization
between $p_z^{\mathrm{A}}$ orbitals and $\mathrm{Ni}_{3z²-r²}$ is
observed, which gives rise to an unoccupied antibonding state
$\sigma^*$ and two occupied bonding states $\sigma_1$ and $\sigma_2$.
The LDA gaps for $\sigma^*-\sigma_1$ and $\sigma^*-\sigma_2$ are 2 eV
and 4 eV, respectively. On the other hand, the $p_z^{\mathrm{B}}$
orbitals (occupied in the spin up channel and unoccupied in the spin
down one) hardly interact with Ni $d$ states and therefore remain
unmodified.  The vdw-DF band structures (evaluated at the vdw-DF
relaxed distance of 3.50 \AA), on the other hand, resemble the free
graphene, preserving the Dirac point and only shifting it up by $0.13
\,$ eV.  A very similar behaviour is found for Co and Pd.  For the
remaining interfaces both the LDA and vdw-DF bandstructures resemble
that of free standing graphene with the Dirac point shifted with
respect to the metal Fermi level. The Fermi level shifts and
calculated charge transfer between the metal and graphene are
summarized in Table \ref{tab:table1}.

Since LDA is known to underestimate band gaps significantly we have
also performed G$_0$W$_0$ calculations for the graphene-Ni structures
corresponding to the LDA and vdw-DF distances.\cite{gw_details} In
both cases we find no noteworthy difference between the G$_0$W$_0$ and
DFT Kohn-Sham band structures close to the $K$-point.

Recent experimental work on the graphene/Ni interface is based on
Angle Resolved Photoemission Spectroscopy (ARPES). The ARPES band
structures reveal a band gap at the graphene $K$-point thus suggesting
some hybridization between the graphene and Ni
orbitals\cite{Varykhalov:2009,Gruneis:2008}. Earlier LEED measurements
found a Ni-graphene bond distance of $~2.1$\AA\cite{LEED}.  We note
that both of these results are in line with the LDA calculations. On
the other hand LDA is not expected to work well for highly
inhomogeneous systems such as the interface structures investigated
here. These results indicate a difficulty for the present vdw-DF in
describing systems with mixed bonding character, in line with the
conclusions of \cite{Romaner:2009}.

Fig.\ \ref{fig:binding_contributions} shows the total (full lines) and
the exchange-only (dashed lines) binding energy curves for revPBE, PBE
and LDA in the case of graphene on Cu(111) (left panel) and on Ni(111)
(right panel). The exchange-only energies are calculated without
including the correlation energy term in the exchange-correlation
functional and have been evaluated non selfconsistently. Clearly, the
bonding for the physisorbed graphene on Al, Cu, Ag, Au and Pt
originates partially from the exchange term in the LDA xc-functional,
as shown for Cu as an example in the left panel of Fig.\
\ref{fig:binding_contributions} . This is in principle incorrect since
van der Waals interactions are purely non-local correlation effects.
The weak bonding predicted by the vdw-DF functional, similar in
magnitude to the LDA results, is produced by the correlation term
instead, which is physically correct. Interestingly, this applies to
the Ni/graphene system as well, as shown in the right panel of Fig.\
\ref{fig:binding_contributions}..  The LDA exchange-only curve shows a
broad and weak attractive contribution between $2 \, \mathrm{\AA}$ and
$5 \, \mathrm{\AA}$ with two local minima.  We note that in a genuine
covalent bond the exchange contribution to the binding energy is
generally significantly larger than in this case.  The revPBE and PBE
exchange-only curves are repulsive at all separations for both
systems. This is the typical behaviour which is observed in van der
Waals bonded dimers or organic molecules on
surfaces\cite{Dion:2004,Moses:2009}

In conclusion we have performed DFT calculations of graphene adsorbed
on different metal surfaces using the recently developed vdw-DF
functional which explicitly includes non-local correlations. For Al,
Cu, Ag, Au and Pt both LDA and vdw-DF consistently predicts a weak
binding. Similar weak binding is found for Ni, Co and Pd with the
vdw-DF where LDA on the other hand predicts stronger binding and
significant hybridization between graphene and metal $d$-states. At
the vdw-DF binding distances graphene's band structure was shown to be
essentially unaffected by the substrate. This appears to be in
conflict with LEED and ARPES measurements for graphene on Ni,
indicating that more work is needed in order to reconcile experiments
and theory for the graphene-metal interface.

\begin{acknowledgments}
  We thank Jens N{\o}rskov and Bengt Lundqvist for useful discussions.
  The authors acknowledge support from the Danish Center for
  Scientific Computing through grant HDW-1103-06. The Center for
  Atomic-scale Materials Design is sponsored by the Lundbeck
  Foundation.
\end{acknowledgments}
\bibliographystyle{apsrev}

\begin{thebibliography}{23}
\expandafter\ifx\csname natexlab\endcsname\relax\def\natexlab#1{#1}\fi
\expandafter\ifx\csname bibnamefont\endcsname\relax
  \def\bibnamefont#1{#1}\fi
\expandafter\ifx\csname bibfnamefont\endcsname\relax
  \def\bibfnamefont#1{#1}\fi
\expandafter\ifx\csname citenamefont\endcsname\relax
  \def\citenamefont#1{#1}\fi
\expandafter\ifx\csname url\endcsname\relax
  \def\url#1{\texttt{#1}}\fi
\expandafter\ifx\csname urlprefix\endcsname\relax\def\urlprefix{URL }\fi
\providecommand{\bibinfo}[2]{#2}
\providecommand{\eprint}[2][]{\url{#2}}

\bibitem{Novoselov:2004}
  K. S. Novoselov, A. K. Geim, S. V. Morozov, D. Jiang, Y. Zhang, S. V. Dubonos, I. V. Grigorieva, A. A. Firsov 
  Science {\bf 306}, 666 (2004).

\bibitem{Geim:2007}
  A. K. Geim, K. S. Novoselov
  Nature Mater. {\bf 6}, 183 (2007).

\bibitem{Wintterlin:2009}
  J. Wintterlin, M.-L. Bocquet
  Surf. Sci. {\bf 603}, 1841 (2009).

\bibitem{Kim:2009}
 Keun S. Kim, Yue Zhao, Houk Jang, Sang Y. Lee, Jong M. Kim, Kwang S. Kim, Jong-Hyun Ahn, Philip Kim, Jae-Young Choi, Byung H. Hong
   Nature {\bf 457}, 706 (2009).

\bibitem{Coraux:2009}
  Johann Coraux, Alpha T N'Diaye, Martin Engler, Carsten Busse, Dirk Wall, Niemma Buckanie, Frank-J Meyer zu Heringdorf, Raoul van Gastel, Bene Poelsema and Thomas Michely
   New J. Phys. {\bf 11}, 039801 (2009).

\bibitem{Sutter:2008}
  Peter W. Sutter, Jan-Ingo Flege and  Eli A. Sutter
  Nature Mater. {\bf 7}, 406 (2008).

\bibitem{Hobza:1995}
  P. Hobza, J. \v{S}poner and T. Reschel 
  J, Comput. Chem. {\bf 16}, 1315 (1995).

\bibitem{Kristyan:1994}
  S. Kristyan, P. Pulay
  Chem. Phys. Lett. {\bf 229}, 175 (1994).

\bibitem{Ruiz:1995}
  E. Ruiz, D. R. Salahub, A. Vela
  J. Am. Chem. Soc. {\bf 117}, 1141 (1995).

\bibitem{Giovannetti:2008}
  G. Giovannetti, P. A. Khomyakov, G. Brocks, V. M. Karpan, J. van den Brink and P. J. Kelly,
  Phys. Rev. Lett. {\bf 101}, 026803 (2008)

\bibitem{Zhang:1998}
  Y. K. Zhang and W. T. Yang
  Phys. Rev. Lett. {\bf 80}, 890 (1998).

\bibitem{PBE:1996}
  J. P. Perdew, K. Burke, M. Ernzerhof
  Phys. Rev. Lett. {\bf 77}, 3865 (1996)

\bibitem{Fuentes:2008}
  M. Fuentes-Cabrera, M.I. Baskes, A.V. Melechko, M.L. Simpson,
  Phys. Rev. B {\bf 77}, 035405 (2008)

\bibitem{Dion:2004}
  M. Dion, H. Rydberg, E. Schr\"{o}der, D. C. Langreth and B. I. Lundqvist,
  Phys. Rev. Lett. {\bf 92}, 246401 (2004)

\bibitem{Dion:2005}
  M. Dion, H. Rydberg, E. Schr\"{o}der, D. C. Langreth and B. I. Lundqvist,
  Phys. Rev. Lett. {\bf 95}, 109902 (2005)

\bibitem{Puzder:2006}
  A. Puzder, M. Dion and D. C. Langreth,
  J. Chem. Phys. {\bf 124}, 164105 (2006)

\bibitem{Thonhauser:2006}
  T. Thonhauser,  A. Puzder and D. C. Langreth,
  J. Chem. Phys. {\bf 124}, 164106 (2006)

\bibitem{Ziambaras:2007}
  E. Ziambaras, J. Kleis, E. Schr\"{o}der, and P. Hyldgaard
  Phys. Rev. B {\bf 76}, 155425 (2007).

\bibitem{Kleis:2007}
  J. Kleis, B.I. Lundqvist, D.C. Langreth, and E. Schr\"{o}der,
  Phys. Rev. B {\bf 76}, 100201 (2007)

\bibitem{Cooper:2008}
  V.R. Cooper, T. Thonhauser, A. Puzder, E. Schr\"{o}der, B.I. Lundqvist, and D.C. Langreth,
  J. Am. Chem. Soc. {\bf 130}, 1304 (2008).

\bibitem{Langreth:2009}
   D. C. Langreth, B. I. Lundqvist, S. D. Chakarova-K\'{a}ck, V. R. Cooper, M. Dion, P. Hyldgaard, A. Kelkkanen, J. Kleis, Lingzhu Kong, Shen Li, P. G. Moses, E. Murray, A. Puzder, H. Rydberg, E. Schr\'{o}der and T. Thonhauser
   J. Phys.: Condens. Matter {\bf 21}, 084203 (2009).

\bibitem{Moses:2009}
  P.G. Moses, J.J Mortensen, B.I. Lundqvist, J.K. N\o rskov,
  J. Chem. Phys. {\bf 130}, 104709 (2009).

\bibitem{Romaner:2009}
   L. Romaner, D. Nabok, P. Puschnig, E. Zojer, C. Ambrosch-Draxl
   New J. Phys. {\bf 11}, 053010 (2009).

\bibitem{Soler:2009}
  G. Roman- Perez, J.M Soler,
  Phys. Rev. Lett. {\bf 103} 096102 (2009).

\bibitem{gw_details} The G$_0$W$_0$ calculations were performed with
   the Yambo code\cite{yambo}. We included 100 empty bands (corresponding to 40 eV
   above the Fermi level), a 12x12 $k$-point for the simple rhombohedral
   unit cell, and a plasmon frequency of 1 Hartree.

\bibitem{yambo}
A. Marini, C. Hogan, M. Grüning, D. Varsano, 
Computer Physics Communications {\bf 180}, 1392 (2009).

\bibitem{Varykhalov:2009}
  A. Varykhalov, J. Sanchez-Barriga, A. M. Shikin, C. Biswas, E. Vescovo, A. Rybkin, D. Marchenko, O. Rader
  Phys. Rev. Lett {\bf 101}, 157601 (2008).

\bibitem{Gruneis:2008}
  A. Gr\"{u}neis, D. V. Vyalikh
  Phys. Rev. B {\bf 77}, 193401 (2008).

\bibitem{LEED}
Y. Gamo, A. Nagashima, M. Wakabayashi, M. Terai, and C. Oshima,
Surf. Sci. 374, {\bf 61} (1997).

\bibitem{Mortensen:2005}
  J. J. Mortensen, L. B Hansen, and K. W. Jacobsen,
  Phys. Rev. B {\bf 71}, 035109 (2005)

\bibitem{Tang:2009}
  W. Tang, E. Sanville, and G. Henkelman
  J. Phys.: Condens. Matter {\bf 21}, 084204 (2009).
    
\end{thebibliography}

\end{document}